# Highly Efficient Deep UV Generation by Four-Wave Mixing in Gas-Filled Hollow Core Photonic Crystal Fiber

Federico Belli[1,2]*, Amir Abdolvand[1], John C. Travers[1,2] and Philip St.J. Russell[1,3]

[1] *Max Planck Institute for the Science of Light, Staudtstrasse 2, 91058 Erlangen, Germany*
[2] *School of Engineering and Physical Sciences, Heriot-Watt University, Edinburgh EH14 4AS, United Kingdom*
[3] *Department of Physics, Friedrich-Alexander Universität, 91058 Erlangen, Germany*
*Corresponding author: f.belli@hw.ac.uk*

**We report on a highly-efficient experimental scheme for the generation of deep-ultraviolet ultrashort light pulses using four-wave mixing in gas-filled kagomé-style photonic crystal fiber. By pumping with ultrashort, few µJ, pulses centered at 400 nm, we generate an idler pulse at 266 nm, and amplify a seeded signal at 800 nm. We achieve remarkably high pump-to-idler energy conversion efficiencies of up to 38%. Although the pump and seed pulse durations are ~100 fs, the generated ultraviolet spectral bandwidths support sub-15 fs pulses. These can be further extended to support few-cycle pulses. Four-wave mixing in gas-filled hollow-core fibres can be scaled to high average powers and different spectral regions such as the vacuum ultraviolet (100-200 nm).**

Wavelength conversion of laser light in nonlinear crystals represents the foundation of nonlinear optics, from the very early days of lasers [1] to more recent broadband optical parametric amplifiers (OPAs) [2] and pulse characterization techniques [3–5]. However, efficient up-conversion schemes based on nonlinear crystals are inherently limited to the near ultraviolet (UV) because of two main bottlenecks: (i) the limited transmission range, damage and two-photon ionization threshold of conventional nonlinear crystals, and (ii) the large group velocity mismatch (GVM) arising from crystal dispersion, especially when pumping in the UV, or with far-detuned spectral components. Non-collinear geometries in bulk nonlinear crystals [2,6] have provided a viable solution from the mid-infrared (IR) to the near UV (NUV, λ > 300 nm). However, extending these techniques further to the deep UV (DUV) or even to the vacuum UV (VUV, λ < 200 nm) is extremely challenging and generally results in low conversion efficiencies (≪ 1%) and often involves complex crystal growth processes [6,7].

An alternative route is provided by strong-field nonlinear optics [8], such as high-harmonic generation or non-collinear four-wave mixing (FWM) in bulk gases [9,10]. Light atomic or molecular gases provide smoother dispersion profiles down to the VUV, but their inherently low nonlinearities require high peak powers (GW) for frequency conversion in these systems, leading therefore to poor or highly demanding average power scalability.

In this letter, we propose and report for the first time a four-wave mixing scheme in broadband-guiding gas-filled kagomé-style photonic crystal fiber (kagomé-PCF). These fibers, as do the related single-ring PCFs, guide light by anti-resonant reflection. We demonstrate phase-matched FWM processes for generating UV radiation with broad bandwidths and low pump pulse energy requirements—made possible by tight field confinement over extended interaction lengths. In this way we demonstrate a simple and viable solution to the complex problem of average power scaling of ultrashort pulses in the deep UV, while demonstrating at the same time increased conversion efficiencies and broader spectral bandwidths.

A large number of nonlinear effects have been predicted and observed in gas-filled hollow-core fibers, examples being soliton self-compression [11], dispersive-wave (DW) emission [12,13] and frequency wiggling [14] as well as tunable high-harmonic generation [15]. In this context, visible to vacuum ultraviolet (VUV) generation has mainly been through DW emission from a self-compressing pulse in monatomic gases [13] as well as VUV supercontinuum generation in molecular gases [16]. DW emission has enabled generation of wavelength-tunable vacuum ultraviolet light with efficiencies ranging from 1% at 120 nm to 10% at 550 nm [16,17], from an ultrashort pump pulse centered at 800 nm, and it is already finding applications in angle [18] and time resolved photoemission spectroscopy [19]. One disadvantage of

DW emission is that the emitted spectra are very tightly coupled to the soliton propagation dynamics in gas-filled kagomé-PCF, which severely limits independent control over the bandwidth and the central emission frequency. Moreover, the driving pump pulse needs to have a short temporal duration (sub 30 fs in the near-infrared) for efficient VUV generation [12].

Ultrafast FWM does not suffer these restraints while offering higher conversion efficiencies.

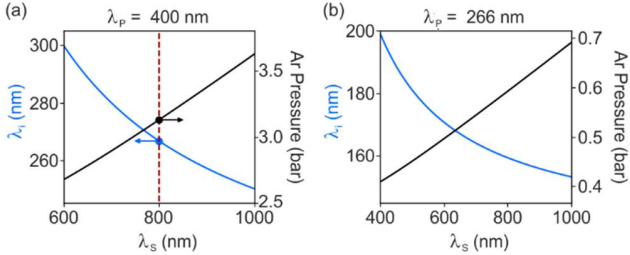

Fig. 1: Deep (a) and vacuum (b) ultraviolet generation schemes. Idler wavelengths ($\lambda_i$, blue lines) as a function of the seeded signal wavelengths ($\lambda_s$) for the given pump beams centred at 400 nm (a) and 266 nm (b), along with corresponding argon pressures (black lines) required to achieve linear phase-matching $\Delta\beta(\omega_p, \omega_i) = 0$ in a kagomé-PCF with a core diameter of 26 μm. The red dashed line marks the wavelength position of the seed signal used in our experiments.

FWM processes are based on third-order nonlinearity, and in the following we will consider a *degenerate seeded* scheme, in which the two pump beams, considered here to be degenerate and originating from the same pump pulse, are centered at an angular frequency $\omega_p$ and are down-converted to a signal beam with angular frequency $\omega_s$ and upconverted to an idler beam centred at $\omega_i = 2\omega_p - \omega_s$. The linear wavevector mismatch is $\Delta\beta(\omega_p, \omega_s) = \beta(\omega_s) + \beta(\omega_i) - 2\beta(\omega_p)$, where $\beta(\omega)$ is the propagation constant calculated by considering the full-dispersion profile of fundamental HE$_{11}$ mode of a gas-filled hollow glass capillary [12,20]. In kagomé-PCFs filled with gases, the dispersion is widely tunable as a result of the balance between the normal gas dispersion and anomalous waveguide dispersion [12]. By changing the core size, gas species, gas pressure, pump and seed frequencies the linear phase-matching condition can be easily fulfilled over multiple octaves. Two examples of deep-UV (Fig. 1a) and VUV (Fig. 1b) generation for a given core diameter and gas species are shown in Fig. 1. More generally, it is possible to phase-match idler and signal frequencies from the VUV to the mid-infrared, while maintaining guidance in the kagomé-PCF [16]. This is a central characteristic of our system, in contrast to conventional FWM process implemented in bulk nonlinear materials, such as gases and crystals, or in waveguides such as step-index silica fiber [21] or solid-core PCF [22].

Since the linear phase-matching condition neglects several key effects such as pump depletion and self- and cross-phase modulation [23,24], it does not provide an exact match to the optimal energy conversion efficiencies obtained in the experiments. Nevertheless, it provides simple means, independent of both pump and seed peak powers, of illustrating the broad tunability of the system while providing a useful first order approximation to the optimal phase-matching pressures which, as we report, depend on several factors such as the time delay between the pump and idler pulses, as well as the pump and idler peak powers.

The generation of a signal beam via FWM has been previously investigated with ps pulses in hollow-core photonic band-gap PCF by carefully tuning the locations of three transmission windows to coincide with the required pump, signal and idler wavelengths [25]. However, the narrow guidance window and large dispersion oscillation in PBF limit the bandwidth of the signals, and hence the achievable minimum pulse duration and spectral coverage. On the other hand in hollow capillary fibers (bore tube), ultrafast FWM has been used at 100 μJ pump-energy scales, to produce a tunable VUV pulse source [26], with broad bandwidths at low repetition rates (1 kHz). However, average power scaling in hollow capillary fibers is more challenging as a result of the high single-pulse pump-energy requirement. Here, we show that this pump-energy requirement can be drastically reduced by using gas-filled ARFs, thus enabling highly-efficient, watt-level tunable sources in ultraviolet and beyond.

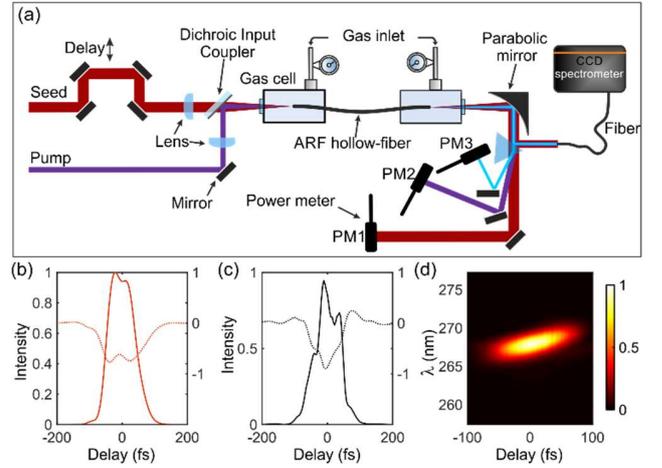

Fig. 2: Layout of the experimental setup (a). Retrieved pulse intensity profile (continuous line) and phase (dashed line) for the seed (red in b) and pump (black in c), together with the experimental X-FROG trace for the pump pulse retrieval (d).

Figure 2 shows the experimental setup, chosen for convenient experimental demonstration. Light from an amplified 1 kHz Ti:sapphire laser was used to generate both a pump pulse at 400 nm (via second harmonic generation (SHG) in a 100 μm BBO crystal), and a co-polarized seed (signal) beam centered at 800 nm with a bandwidth of 5 nm. Using two achromatic lenses and a dichroic mirror, the pump and seed pulses were combined and coupled into the fundamental mode of a kagomé-PCF with core diameter of 26 μm, placed between two gas cells filled with Ar. For these pump and signal wavelengths, the idler is generated at 266 nm. Tuning the pump or signal wavelength, along with the gas pressure, enables tunable DUV generation. In order to precisely control the gas-filling pressure in the system two pressure regulators (flow regulator and back-pressure regulator) were used and a precision motorized delay stage the relative pump-signal delay to be tuned with femtosecond resolution.

Temporal characterization of the 800 nm seed pulse at the fiber input was carried out using a commercial (Grenouille) SHG frequency-resolved optical gating device (FROG), while the 400 nm pump pulse was characterized using a (homebuilt) cross-correlation FROG (X-FROG), consisting of a 100 μm thick BBO crystal for sum-frequency generation with a 40 fs reference pulse at

800 nm. A typical X-FROG trace obtained by scanning the relative pump-reference delay is shown in Fig. 2, together with the retrieved pulse shape for the seed (red) and pump pulses (black). The pump pulse had a duration of ~90 fs and the seed pulse a duration of 95 fs.

The divergent pump, seed and idler beams emerging from the fiber output were collimated with a UV-enhanced aluminum off-axis parabolic mirror, after passing through a 2 mm thick $MgF_2$ window. Reflections from an uncoated $CaF_2$ prism were used to carry out spectral measurements with a calibrated fiber-based CCD spectrometer (Ocean Optics), while the spatially-dispersed beam from the prism was redirected to three different calibrated photodiodes (PM1-3 in Fig. 1 a, Ophir) in order to measure separately the energy in the pump, seed and idler pulses. All the different beam paths were carefully calibrated so as to map the photodiode energy readings to the energies at the fiber output in the three spectral bands corresponding to the pump (350-450 nm), signal (700-900 nm) and idler (240-290 nm). The photodiode readings, together with simultaneous measurements of the relative spectral energy density (SED) from the spectrometer, allowed the energy in the three bands to be accurately estimated by numerical integration across the spectrum and rescaling to the energy measured by the photodiodes.

Fig. 3(a) shows the measured input/output energy for the pump, idler and signal beams, plotted as function of input pump power at an argon pressure of 3.18 bar (determined to be near optimum from a gas pressure scan, as discussed later), when seeding the signal at 800 nm with 250 nJ. The pump pulse at 400 nm is efficiently converted to the idler (blue line) in the deep UV, while amplifying the signal (red line). This is also clear from the pump depletion, idler generation and signal amplification shown in Fig. 3(b). In contrast, negligible idler and signal powers are generated in the unseeded case (see Fig. 3 (c-e)).

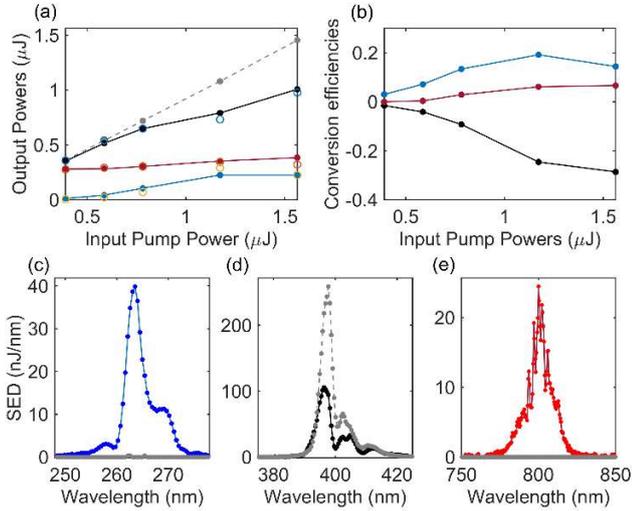

Fig. 3: Experimental pump energy scan with argon at 3.18 bar. (a) Output pump (black), signal (red) and idler (blue) pulse energies as a function of input pump energy, when seeding with a 250 nJ signal pulse. In the absence of seeding the pump energy follows the black dashed line, and the idler and signal power are negligible. (b) Energy conversion efficiencies for idler generation $E_i^{out}$ (blue), signal amplification $(E_s^{out} - E_s^{in})/E_p^{in}$ (red) and pump depletion $(E_p^{out} - E_p^{in})/E_p^{in}$ (black). The dots represent the calibrated energy meter readings, while open circles represent the estimated energies from the calibrated spectrometer. (c-e): Output spectral energy densities (nJ/nm) for an input pump energy of 1.2 μJ when seeding with 250 nJ signal seed (blue (c), black (d) and red (e)) and without signal seed (gray), for the idler (c), pump (d) and signal (e).

Moreover, since the FWM process is based on non-resonant Kerr nonlinearities, energy and photon number are conserved, leading to $(E_p^{in} - E_p^{out})/\omega_p = (E_s^{out} - E_s^{in})/\omega_s + E_i^{out}(z)/\omega_i$ where the $E_n$ are the energies of the interacting fields. As shown in Fig. 3(b), the depleted pump power (black) is converted to the idler (blue) and signal (red) in a fixed proportion of 3:1. This is the result of conversion of two pump photons, frequency $\omega_p = 2\omega_0$, to single photons at frequencies $\omega_s = \omega_0$ and $\omega_i = 3\omega_0$, where $\nu_0 = (\omega_0/2\pi) = 375$ THz is the Ti:sapphire laser frequency at 800 nm.

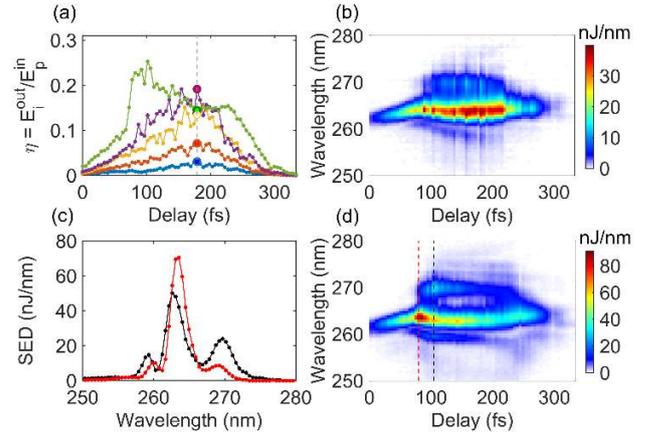

Fig. 4: Pump-seed delay scan and input power scan. (a) Conversion efficiencies $\eta = E_i^{out}/E_p^{in}$ as function of the relative delay between the pump and seed pulse (with arbitrary offset) for 3.18 bar gas pressure and different power in the pump: 0.4 μJ (blue), 0.6 μJ (red), 0.8 μJ (yellow), 1.2 μJ (purple) and 1.6 μJ (green). Vertical line marks the temporal position of the power meter readings, that have been used to rescale the relative energy measurement obtained from the spectrometer at every delay position. (b,d): Estimated spectral energy density for fixed input pump power of 1.2 μJ (b) and 1.5 μJ (d). (c) Estimated spectral energy density for the two fixed pump-seed delays marked with lines in (d).

The complex interactions between signal, pump and idler pulses mean that the efficiency of the idler generation depends on many parameters: the pump and seed powers, the pump to seed delay $\tau_{ps}$, the gas-filling pressure and the fiber parameters. These must all be carefully chosen to optimize the FWM process.

At the highest pump energy, ~1.5 μJ in Fig. 3 (a-b), the idler conversion efficiency drops and seed amplification saturates. This is the consequence of power-dependent changes in nonlinear phase-matching, and the effects of the pump-seed delay. The results in Fig. 3 were obtained for a fixed pump-seed delay and at a fixed gas-pressure. In Fig. 4 we show a full scan of the pump-seed delay for the same experimental conditions as in Fig. 3.

Since both the pump and the signal pulses are 100 fs long, their temporal overlap plays a crucial role in obtaining the optimal conversion efficiency, especially at the highest pump energy. The dependence of the idler conversion efficiency on the relative delay between pump and the seed is shown for different pump powers in Fig. 4 (a). The generated SED is plotted versus pump-seed delay $\tau_{pi}$ in Figs. 4 (b&d) for pump powers of 1.2 μJ and 1.5 μJ. The changes in spectral shape, shown by black and red curves in Fig. 4(c), are the result of pump-driven cross-phase modulation that is directly transferred to the idler beam.

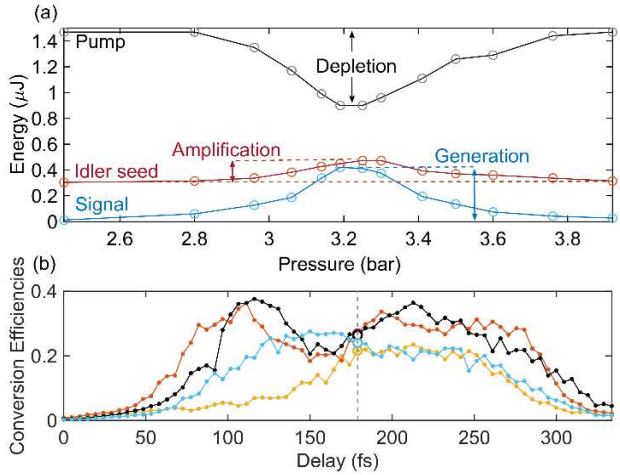

Fig. 5: Pressure and delay scan. (a) Measured energis at the fiber output for pump (black), idler (blue) and signal (red) as function of pressure at a given temporal delay $\tau_{pi}$ (marked with full circles in (b)). Estimated conversion efficiencies as function of pump-seed delay for pressures of 3.14 (cyan), 3.19 (purple), 3.25 (orange) and 3.3 (black) bar. The full circles in (b) mark the temporal position of the photodiode energy measurement. The maximum conversion efficiency estimated from the calibrated spectrometer is 38%.

To determine the maximum DUV conversion efficiency attainable in the system, we varied the argon gas-pressure and the pump-seed delay, while keeing the pump energy fixed at its maximum value of 1.5 μJ. The results are shown in Fig. 5, where the pump (black), signal (red) and idler (blue) energies are plotted as a function of the pressure (Fig. 5(a)) and delay (Fig. 5(b)). When optimized, the system generated 570 nJ DUV pulses centered at ∼265 nm, with a bandwidth capable of supporting transform-limited sub-15 fs pulses, and a conversion efficiency of 38% from the input pump energy.

In summary, we have investigated for the first time the generation of deep-ultraviolet light pulses using FWM in an anti-resonant fiber filled with noble gas. We have achieved conversion efficiencies to the DUV of up to 38% of the pump energy at the fiber input, when weakly seeding the process. Due to relatively modest requirements in terms of pulse energy (∼1 μJ) and pulse duration (∼100 fs), average power scaling to several tens of W can be reasonably expected with current laser technology and the proven average-power handling of anti-resonant kagomé and single-ring hollow-core PCFs. Moreover, due to the extended guidance range in the UV region, ultrafast FWM in hollow-core fibers offers an attractive route to up-conversion to the VUV with high conversion efficiencies.

## References

[1] P. A. Franken, A. E. Hill, C. W. Peters, and G. Weinreich, Phys. Rev. Lett. **7**, 118 (1961).
[2] G. Cerullo and S. De Silvestri, Rev. Sci. Instrum. **74**, 1 (2003).
[3] R. Trebino, *Frequency-Resolved Optical Gating: The Measurement of Ultrashort Laser Pulses* (Springer US : Imprint : Springer, Boston, MA, 2000).
[4] A. Monmayrant, S. Weber, and B. Chatel, J. Phys. B At. Mol. Opt. Phys. **43**, 103001 (2010).
[5] A. Baltuska, M. S. Pshenichnikov, and D. A. Wiersma, IEEE J. Quantum Electron. **35**, 459 (1999).
[6] P. Trabs, F. Noack, A. S. Aleksandrovsky, A. I. Zaitsev, and V. Petrov, Opt. Lett. **41**, 618 (2016).
[7] C. T. Chen, G. L. Wang, X. Y. Wang, and Z. Y. Xu, Appl. Phys. B **97**, 9 (2009).
[8] T. Brabec, editor , *Strong Field Laser Physics* (Springer, New York, NY, 2008).
[9] M. Ghotbi, M. Beutler, and F. Noack, Opt. Lett. **35**, 3492 (2010).
[10] M. Beutler, M. Ghotbi, F. Noack, and I. V. Hertel, Opt. Lett. **35**, 1491 (2010).
[11] P. St.J. Russell, P. Hölzer, W. Chang, A. Abdolvand, and J. C. Travers, Nat. Photonics **8**, 278 (2014).
[12] J. C. Travers, W. Chang, J. Nold, N. Y. Joly, and P. St.J. Russell, JOSA B **28**, A11 (2011).
[13] N. Y. Joly, J. Nold, W. Chang, P. Hölzer, A. Nazarkin, G. K. L. Wong, F. Biancalana, and J. Russell, P. St.J., Phys Rev Lett **106**, 203901 (2011).
[14] F. Belli, A. Abdolvand, J. C. Travers, and P. St.J. Russell, Phys. Rev. A **97**, 013814 (2018).
[15] F. Tani, M. H. Frosz, J. C. Travers, and P. St.J. Russell, Opt. Lett. **42**, 1768 (2017).
[16] F. Belli, A. Abdolvand, W. Chang, J. C. Travers, and P. St.J. Russell, Optica **2**, 292 (2015).
[17] A. Ermolov, K. F. Mak, M. H. Frosz, J. C. Travers, and P. St.J. Russell, Phys Rev A **92**, 033821 (2015).
[18] H. Bromberger, A. Ermolov, F. Belli, H. Liu, F. Calegari, M. Chávez-Cervantes, M. T. Li, C. T. Lin, A. Abdolvand, P. St.J. Russell, A. Cavalleri, J. C. Travers, and I. Gierz, Appl. Phys. Lett. **107**, 091101 (2015).
[19] N. Kotsina, F. Belli, S. Gao, Y. Wang, P. Wang, J. C. Travers, and D. Townsend, J. Phys. Chem. Lett. (2019).
[20] E. A. J. Marcatili and R. A. Schmeltzer, Bell Syst. Tech. J. **43**, 1759 (1964).
[21] J. Hansryd, P. A. Andrekson, M. Westlund, Jie Li, and P.-O. Hedekvist, IEEE J. Sel. Top. Quantum Electron. **8**, 506 (2002).
[22] W. J. Wadsworth, N. Joly, J. C. Knight, T. A. Birks, F. Biancalana, and P. St.J. Russell, Opt. Express **12**, 299 (2004).
[23] G. Cappellini and S. Trillo, JOSA B **8**, 824 (1991).
[24] G. Agrawal, *Nonlinear Fiber Optics*, Fifth Edition (Academic Press, 2013).
[25] S. O. Konorov, A. B. Fedotov, and A. M. Zheltikov, Opt. Lett. **28**, 1448 (2003).
[26] L. Misoguti, S. Backus, C. Durfee, R. Bartels, M. Murnane, and H. Kapteyn, Phys. Rev. Lett. **87**, (2001).
[27] F. Köttig, F. Tani, C. M. Biersach, J. C. Travers, and P. St.J. Russell, Optica **4**, 1272 (2017).